\documentstyle[aps]{revtex}
\begin{document}
\title{Self-consistent 
calculation of 
particle-hole diagrams\\ on 
the Matsubara frequency: 
FLEX approximation}
\author{J.J. 
Rodr\'{\i}guez-N\'u\~nez}
\address{Instituto de 
F\'{\i}sica, 
Universidade Federal 
Fluminense,\\ 
Av.\ Litor\^anea S/N, 
Boa Viagem, \\
24210-340 Niter\'oi RJ, 
Brazil. \\e-m: jjrn@if.uff.br}
\author{S. Schafroth}
\address{Physik--Institut 
der Universit\"at Z\"urich,\\
Winterthurerstrasse 190, 
CH--8057 Z\"urich, Switzerland.}
\date{\today}
\maketitle
\begin{abstract}
We implement the 
numerical method 
of summing Green 
function diagrams 
on the Matsubara 
frequency axis 
for the fluctuation 
exchange (FLEX)  
approximation. Our 
method has previously 
been applied to the 
attractive Hubbard 
model for low 
density. Here we 
apply our numerical 
algorithm to the Hubbard model 
close to half filling ($\rho = 0.40$),  
and for $T/t = 0.03$, 
in order to study the 
dynamics of one- and 
two-particle Green functions.  
For the values of the  
chosen parameters we see the 
formation of three branches 
which we associate 
with the a two-peak structure in the 
imaginary part of the self-energy.  
From the imaginary part of the self-energy 
we conclude that our system is a Fermi liquid 
(for the temperature investigated here),  
since Im$\Sigma(\vec{k},\omega) \approx w^2$ 
around the chemical potential. 
We have  
compared our fully 
self-consistent 
FLEX solutions with 
a lower 
order approximation 
where the internal 
Green functions are 
approximated 
by free Green functions. These two 
approches, i.e., the fully 
selfconsistent and 
the non-selfconsistent 
ones give different 
results for the parameters 
considered here. 
However, they have 
similar global 
results for small densities. 

Pacs numbers: 74.20.-Fg, 74.10.-z,74.60.-w, 74.72.-h 
\end{abstract}


%\begin{multicols}{2}

\section{Introduction}

	High-temperature 
superconductors\cite{B-M} 
 a wide range 
of behaviour 
atypical\cite{issues} of the standard 
band-theory of metals, 
since at half filling 
they should 
be metals while they 
happen to 
be insulators. Then, 
correlations 
are important. This 
leads us to consider 
that the 
theory of strongly correlated 
Fermin systems plays 
an important 
role for the pairing mechanism  
and other properties of the 
cuprates\cite{Dagotto}. 
The current strategy (due to 
Anderson)\cite{Anderson} to address 
the problem of high-$T_c$ superconductivity 
is to try to find a theory accounting 
for the normal state properties of the 
cuprates. This goes in analogy with the 
ordinary BCS theory where firstly the 
normal state is identified to be a Landau 
Fermi liquid metal and then is 
found an electron pairing mechanism (unknown up to 
now in the cuprates) which destabilizes the 
normal state phase towards a superconducting 
state.  
In particular, 
due to their pronounced 
two-dimensional 
character the one-band 
Hubbard Hamiltonian of the 
Cu-3d hole states has 
been taken as 
one of the essential models\cite{stuff}. 
The Hubbard model 
is the simplest many-particle 
model one can write 
down, which cannot be reduced 
to a single-particle 
theory\cite{AA}. 
To study this model, the simultaneous 
evaluation of the frequency and 
momentum dependence of the one- 
and two-particle Green functions 
is crucial. Here, we adopt the 
opposite view of infinite 
dimensions\cite{GKKR,Vollhardt} which 
neglects the important role 
of spatial correlations.  

	The 
Hubbard model has 
been a subject of 
intense numerical 
studies. In 
spite of exact 
diagonalization and 
quantum Monte Carlo 
studies\cite{9,10,11,Morgenstern}, 
these 
methods are 
restricted to indeed 
small finite size 
systems. Another 
route which takes 
both dynamical 
as well as momentum space 
correlation into account is the 
use of diagrammatic 
approaches\cite{AMT}. 
We 
should mention 
the self-consistent 
summation 
of all bubble and 
ladder diagrams, 
so called fluctuation 
exchange approximation, or 
FLEX\cite{12,13,14,15,16},\cite{21}. 
This 
approximation (FLEX) 
is conserving 
in the sense of Baym and 
Kadanoff\cite{BK}, i.e., it is 
consistent with 
microscopic conservation 
laws for particle number, energy, 
and momentum. We recall that the FLEX 
approximation belongs to the class of 
$\Phi$-derivable approximations which have 
been discussed a long time ago by Wortis\cite{Wortis} 
in the context of perturbative series expansion of 
thermodynamic properties of models described by 
lattice Hamiltonians, mainly the Ising model. 
Wortis\cite{Wortis} gives a set of steps to 
construct a $\Phi$-derivable approximation. 

	In the FLEX approximation there is an 
interesting feature in the effective interaction: it 
is temperature and doping dependent through 
the spin susceptibility, $\chi(\vec{q},\omega)$. The 
effective interaction depends on the properties 
of the quasiparticles. One thus has a system in 
which, since the effective interaction both 
modifies the quasi-particle behavior and is itself 
alterated by that changed quasi-particle behavior, 
non-linear feedback, either positive or negative, 
can play a significant role\cite{Pines}. According 
to the FLEX approach, the dominant contribution to 
the magnetic interaction between planar quasi-particles 
is assumed to come from spin-fluctuation exchange, and 
so will be proportional to $\chi(\vec{q},\omega)$. 
In the scheme of Pines and co-workers\cite{Pines}, 
$\chi(\vec{q},\omega)$ is taken from experiments, while 
in the present work we have a fully self-consistent 
close set of equations to solve. 

	Recently, Schmalian, Langer, 
Grabowski and Bennemann\cite{17}  
have performed the  
calculation of the dynamical 
properties on the 
real axis, instead of 
the Matsubara 
frequencies. Here  
we perform our 
calculations on the Matsubara axis 
(1024 frequencies) obtaining  
the dynamical properties 
for all frequency 
range after analytically 
continuing the one- and 
two-particle 
Green functions with Pad\'e 
approximants\cite{Serene-Vidberg}.

	Our method for solving 
numerically the FLEX equations  
is based on the Fast Fourier 
tranformation (FFT)\cite{Dvries} 
which is 
suitable for the evaluation of 
the corresponding integrals. 
We refer to reference\cite{18} 
where this technique is 
documented. In order to reach high 
densities, while working 
on the imaginary time, we have 
started from small densities (or 
low chemical potential). For each 
value of the chemical potential 
we have reached self-consistency 
and then we move in small 
steps to a higher chemical potential. 
We have proceeded this way 
since the next Green function is 
constructed from the nearby 
previous one. So, we 
always stay close to the 
real solution until we reach the 
density we are looking for.

	In Section \ref{eqsmot} we 
describe our model Hamiltonian 
and the FLEX equations to be solved 
on the imaginary time. In 
Section \ref{3}, we present the data 
for the one- and 
two-particle spectral functions, 
the self-energy and density 
of states. It is observed the 
formation of three branches in the 
spectral density. We analyze, following 
the self-consistency of our equations, the 
presence of these two peaks as due to the 
presence of two peaks in the imaginary part 
of the effective interaction.  
We have compared our solutions 
with a low order approximation  
which consists in approximating the 
internal Green functions 
by free ones. Our results show 
that the calculations 
agree globally with each other for 
small densities. The 
dynamical quantities are radically 
different for the density/spin  
investigated in this paper ($\rho = 
0.40$). In Section \ref{4} 
we discuss our result and present 
our conclusions.

\section{The model and the FLEX 
equations.}\label{eqsmot}

The Hubbard Hamiltonian is defined as
\begin{eqnarray}\label{Ham}
    H = - t\sum_{<ll'>\sigma}
    c_{l\sigma}^{\dagger}c_{l'\sigma}
   + U \sum_l n_{l\uparrow}
    n_{l\downarrow}
   - \mu \sum_{l\sigma} 
   n_{l \sigma}~~,
\end{eqnarray}
where the $c_{l\sigma}^{\dagger}
$($c_{l\sigma}$) 
are creation 
(annihilation)
operators for electrons 
with spin $\sigma$.  The 
number-operator 
is $n_{l\sigma}
\equiv c_{l\sigma}^{\dagger}
c_{l\sigma}$, 
$t$ is the hopping matrix element
between nearest neighbours $l$ and 
$l'$, $U$ is the on-site interaction
and $\mu$ is the chemical potential 
in the grand canonical ensemble.
Here we consider a repulsive 
interaction, $U>0$. 

	The one-particle 
Green function 
is expressed in terms of 
the self-energy, 
$\Sigma({\vec{k}},i\omega_n)$, by 
\begin{equation}\label{Dyson}
G({\vec{k}},i\omega_n) = 
\frac{1}{i\omega_n 
- \varepsilon_{\vec{k}} 
- \Sigma({\vec{k}},i\omega_n)} ~~~,
\end{equation}
\noindent
where $\varepsilon_{\vec{k}} =  
-2t(cos(kx) + cos(ky)) - \mu$. 
The self-energy 
of the Hubbard model 
(Eq. (\ref{Ham})) 
within the 
FLEX approximation 
is given as 
follows\cite{12}
\begin{equation}\label{selfenergy}
\Sigma({\vec{k}},i\omega_n) 
= \frac{T}{N} 
\sum_{{\bf q},i\epsilon_m} 
V_{p-h}({\bf q},i\epsilon_m) 
G({\bf k - q},i\omega_n-i\epsilon_m)~~~,
\end{equation} 
\noindent
where the effective 
interaction, 
$V_{p-h}({\bf q},i\epsilon_m)$,  
resulting from the 
summation of 
the bubble and ladder 
diagrams (particle - 
hole channel), is 
given by
\begin{equation}\label{Veff}
V_{p-h}({\bf q},i\epsilon_m) = 
\frac{U^2}{2} 
\chi({\bf q},i\epsilon_m)
\left( \frac{3}
{1-U\chi({\bf q},i\epsilon_m)} 
+ \frac{1}{1 + 
U\chi({\bf q},i\epsilon_m)} 
-2 \right)~~~.
\end{equation}

	Here,
\begin{equation}\label{bubble}
\chi({\bf q},i\epsilon_m) =  
-\frac{T}{N} 
\sum_{{\vec{k}},i\omega_n} 
G({\bf k+q},i\omega_n + 
i\epsilon_m)
G({\vec{k}},i\omega_n) ~~~
\end{equation}
is the particle-hole
 bubble with 
the renormalized 
Green function 
$G({\vec{k}},i\omega_n)$. 
$\omega_n 
= (2n+1) \pi T$ and 
$\epsilon_m 
= 2m\pi T$ are the 
fermionic (odd) 
and bosonic (even) 
Matsubara 
frequencies. Let's 
state that the 
bubble and ladder 
sums represent 
the charge and 
longitudinal and 
the transversal 
spin fluctuations, 
respectively. The 
effect of 
particle-particle 
fluctuations 
(Cooper channel) 
has been left out 
of the present study. 

	The previous 
set of equations 
closes with the 
expression for the 
$\rho$ given by 

\begin{equation}\label{rho}
\rho(T,\mu) = \frac{1}{2} + 
\frac{T}{N} \sum_{{\vec{k}} 
i\omega_n}G({\vec{k}} i\omega_n) ~~~ .
\end{equation}
	The form of Eq. 
(\ref{rho}) 
is well suited for 
numerical 
calculations. It has been 
justified by Schafroth, 
Rodr\'{\i}guez-N\'u\~nez 
and Beck\cite{18}. 
Eq. (\ref{rho}) 
represents a 
generalization of 
Mahan's Eq. 
(3.1.2)\cite{Mahan}. 
Eqs. (\ref{Dyson}-\ref{rho}) 
constitute the 
standard set of FLEX equations 
which have 
to be solved selfconsistently.
 
	In order to 
obtain results 
which are independent 
of finite size,
one should use at 
least some $10^3$
Matsubara frequencies and a grid 
of $30\times 30$ 
lattice points. The above
scheme works in 
principle but a closer
look at the 
equations for $\chi$ 
and $\Sigma$ shows 
that the straight
forward implementation of these 
equations does not 
work in practice. This is
due to the 4--fold 
loops which would 
occur in the computer 
program. Suppose
we use 2000 Matsubara 
frequencies and 
a $30\times 30$ grid. 
Then we have to
carry out for every 
grid point and 
every frequency the 
double sum over all
frequencies and all grid points. 
This are of the order
$(30^2 \times 2000)^2 
= 3.24\times 
10^{12}$ complex 
operations. Even a typical  
super-computer would 
need one to several 
hours to make one
iteration step.

	Since the 
frequency and momentum 
summations are convolutions,
we evaluate them using 
the Fast Fourier 
Transform (FFT). In 
order to do that 
we need to transform 
the FLEX equations 
to direct space-time. 
For example, 
Eqs. (\ref{bubble},\ref{selfenergy}) 
adopt the following forms

\begin{equation}\label{eq1}
\chi({\bf x},\tau) = - 
G({\bf x},\tau) 
G({\bf x},-\tau) ~~~,
\end{equation}
\noindent and 
\begin{equation}\label{eq2}
\Sigma({\bf x},\tau) 
= V_{p-h}({\bf x},\tau) 
G({\bf x},\tau) ~~~.
\end{equation}

	In order be  
self-consistent, 
we calculate the 
frequency integrals 
on the imaginary  
time following a 
different approch 
to the one of 
Schmalian et al\cite{17}. 
According to our 
experience, it is much 
better to work with 
Matsubara frequencies 
due to the fact that we have 
to handle 
neither Fermi nor Bose 
distribution functions. 
As we have previously said in the 
Introduction we have reached 
high densities 
by moving from a selfconsistent 
solution to another close by. 
In other words, we 
do not need going to real 
frequencies 
to have stability in our 
algorithm. 
In Section \ref{3} we 
present our 
results, making special 
emphasis on the 
the way we have reached high densities 
(see Fig. 1).

\section{Numerical results and interpretation}\label{3}

	In Fig. 1 we present the 
density/spin, 
$\rho$, as function of chemical 
potential, $\mu$, for $U/t = 
4.0$ and a 
temperature of $T/t = 0.03$. 
(This temperature 
is slightly above the 
temperature used in 
Ref.\cite{17} of Schmalian et al. 
We have chosen this 
temperature since we can recover 
the asymptotic behavior, i.e., 
$\pi * 512 * T 
= 45t \approx 11.1U$. For 
$T/t < 0.03$ we need to 
increase the number of 
Matsubara frequencies 
(we would need to 
go to RAM memories of 
100 Mb or higher) 
in order to reach the 
asymptotic behavior. 
This asymptotic behavior is 
needed in order to compute 
the Fast Fourier transforms. 
In this figure 
we also show the the equivalent 
results for a low 
order approximation 
which consists in 
replacing in Eqs. 
(\ref{selfenergy}-\ref{bubble}) the 
full one-particle Green function, 
$G({\vec{k}},i\omega_n)$, 
by the one-particle 
free Green function, 
$G_o({\vec{k}},i\omega_n) 
= 1/(i\omega_n - 
\varepsilon_{\vec{k}})$. 
This approximation 
was proposed by Serene 
and Hess\cite{Serene-Hess}. 
We observe 
that the two curves 
almost coincide for 
low densities. For 
high densities, 
this is not the 
case, as we will see 
shortly. At this point 
we would like to 
emphasize that the way we 
have reached high 
densities settles the 
apparent drawbacks 
of the calculations of 
the $T$-matrix approach 
which is a lower order 
approximation to FLEX. There is 
recent calculation by 
Kagan et al\cite{Maxim} where 
they have performed analytical 
calculations for the negative 
Hubbard model in the frame of the 
$T-$Matrix approximation. 

	In Fig. 2 we present 
the spectral 
density, $A(\vec{k},\omega)$, 
as function of 
frequency, $\omega$ for different 
momenta along the 
diagonal of the 
Brillouin zone $(k,k) = 
\pi/16(n,n)$, 
with $n$ an integer. 
We have the 
parameters $U/t = 4.0$, $T/t = 0.03$, 
$\rho = 0.40$. We are 
working with 
a system of $32\times 
32 \times 1024$. 
The function 
$A({\vec{k}},\omega)$ 
is obtained from 
the one-particle Green 
function by means 
of the analytical continuation,
\begin{equation}\label{A}
A({\vec{k}},\omega) 
\equiv - \frac{1}{\pi} 
\lim_{\delta \rightarrow +0} 
Im [G({\vec{k}},\omega 
+ i\delta)] ~~~ .
\end{equation}
	In Eq. (\ref{A}), 
we have chosen 
$\delta/t = 0.01$. After 
selfconsistency, the 
chemical potential turns out 
to be $\mu = 
-~0.95$. We should like to 
point out that the 
chemical potential is 
independent of $\delta$. 
This value is used at the end of 
the calculations 
to perform the analytical 
continuation as done 
by Vidberg and 
Serene\cite{Serene-Vidberg}. 
See also Ref.\cite{17} 
for more details. 
From Fig. 2 we see that 
the spectral functions have 
a well defined peak for small 
momenta and a satellite for 
positive frequency. For large momenta, we 
observe a kind of {\it incoherent peak} for 
positive frequencies and a satellite for 
negative frequencies. This {\it incoherent} 
peak could be decomposed into two peaks, one 
for small positive frequencies and another one 
for large positive frequencies. These results are 
different with respect to the negative Hubbard 
model at small density (T-Matrix Approximation), 
since in the T-Matrix approximation we find two 
peaks in the one-particle spectral function, 
while here we are resolving three peaks, instead.   
The relevance 
of the spectral functions 
in the Hubbard 
model both above and below 
the critical 
temperature, $T_c$ of the 
superconducting 
state has been treated by 
Dahm and 
Tewordt\cite{21} (See their 
Figs. 17). They have compared their 
spectra with photoemission data 
of $La_{2-x}Sr_xCuO4$.
 
	For comparison, 
in Fig. 3, we 
present the spectral 
densities for the 
low order approximation. 
We observe that the peak 
structure is more 
pronunced than in the 
interacting case. This 
effect is also seen in 
the self-energy which 
we show next.

	In Fig. 4 we 
display the imaginary 
part of the self-energy, 
$-Im[\Sigma(n,n,\omega)]$, 
along the diagonal of 
the Brillouin zone. 
The parameters are the 
same as in Fig. 2. We 
observe that energy 
dependence of 
$-Im[\Sigma(n,n,\omega)]$, 
close to the chemical 
potential (the chemical potential is 
located at $\omega = 0$), 
has a gap-like 
structure. $-Im[\Sigma(n,n,\omega)]$ has 
a well defined gap structure close to 
the chemical potential ($w \approx 0$). 
We also see two almost symmetric peaks for 
every moment $(k,k)$. This presents a 
difference with respect to the attractive Hubbard 
model at small densities. For the attractive 
Hubbard model at small densities, there was a 
single peak in frequency for all values of 
$(k,k)$. This single peak structure in 
$-Im[\Sigma(n,n,\omega)]$ was fitted by a two-pole 
Ansatz for the one-particle Green function. From 
Fig. 4 we conclude that, for the studied temperature, 
we have a Fermi liquid system, since 
Im$\Sigma(\vec{k},\omega) \approx \omega^2$ for $\omega$ 
close to the chemical potential. We agree with 
Wermbtex\cite{meanFlex} even when his calculation is 
mean-field like. See the discussion of Fig. 5.

	In Fig. 5 (a) we present the same 
quantity, $-Im[\Sigma(n,n,\omega)]$ 
vs $\omega$, for the low 
order approximation. 
We inmediately 
see a well defined gap 
for high momenta. 
In consequence, the effect 
of fluctuations is to 
reduce the exaggerated gap of the 
non-selfconsistent solution. 
Replacing our internal 
lines by free Green functions is  
equivalent to stay close 
to mean-field treatments, as 
it has been done by 
Kampf and Schrieffer\cite{20}. 
These authors find 
well defined gaps in the 
spectral density which 
are most likely due to 
their mean field treatment. 
The approach mentioned 
in Ref.\cite{20} has been 
critically studied by 
Monthoux\cite{Monthoux}. In Fig. 5 (b) we show 
$-Im[\Sigma(n,0,\omega)]$ vs $\omega$. Here the 
gap is present for all the momenta. 

	As it has been 
said in the 
Introduction the effective 
potential is strongly 
momentum and frequency 
dependent. We 
show in Figs. 6 and 7 the 
imaginary and real 
parts of the effective 
potential, 
respectively, for 
the same set  of parameters 
as given in 
Fig. 2 along the diagonal 
of the Brillouin 
zone. We observe 
that $Im[V_{p-h}(m,m,\omega)]$ 
($Re[V_{p-h}(m,m,\omega)]$) 
is basically odd (even) in 
frequency around the 
chemical potential. This 
symmetry around zero is due to the 
fact that the Hartree 
shift (HS) has been 
substracted from the 
effective interaction. 
The real part of the 
effective interaction, i.e., 
$Re[V_{p-h}(m,0,\omega)]$ 
vs $\omega$ compares 
qualitately well with Fig. 
19(a) of Ref.\cite{21}. 
The authors of Ref.\cite{21} 
have plotted  
the irreducible spin 
susceptibility, 
$\chi(\vec{k},\omega)$ vs $\omega$ for 
some values of $\vec{k}$.

	Now, we are in a position to explain 
the double peak structure in the imaginary 
part of the self-energy. Going to real 
frequencies, Eq. (\ref{selfenergy}) can be rewritten as
\begin{eqnarray}\label{realw}
\Sigma(\vec{k},z) = \frac{1}{N} 
\sum_{\vec{q}}\left[ \int_{-\infty}^{+\infty} 
d\omega V_{p-h}({\bf q},z-\omega)A(\vec{k}-\vec{q},\omega) 
n_F(\omega) \right. \nonumber \\
- \left. \int_{-\infty}^{+\infty} \frac{d\omega}{\pi} 
Im[V_{p-h}(\vec{q},\omega)]n_B(\omega) 
G(\vec{q}-\vec{k},-\omega+z) \right] ~~~~ ,
\end{eqnarray}
where $n_F(\omega)$ and $n_B(\omega)$ are the 
Fermi and Bose distribution functions, respectively.	

	As the peaks of $Im[V_{p-h}(\vec{q},\omega)]$ are 
symmetric in energies, then both terms of 
Eq. (\ref{realw}) contribute.  
As $Im[V_{p-h}(\vec{q},\omega)]$ is 
antisymmetrical for frequencies close to $\mu$ where 
$A(\vec{k},\omega)$ has a peak, then we have two 
contributions to $Im[\Sigma(\vec{k},\omega)]$. Indeed, 
the numerical results for any $(k,k)$ of the self-energy 
mirror the behavior of $V_{p-h}(\vec{q},\omega)$. 
Thus, the two-particle spectrum is introduced into the 
one-particle quantities like $G(\vec{k},\omega)$ and 
$A(\vec{k},\omega)$. 

\section{Conclusions.}\label{4}
	The FLEX approximation, a 
Baym-Kadanoff generalization 
of Hartree-Fock 
theory, has been 
implemented to study 
the frequency and 
momentum dependence 
of the one- and 
two-particle correlation 
functions. We have found the 
existence of 
three peaks in the 
one-particle spectral 
density, for the set of 
parameters investigated 
here. The presence of 
this  
structure in the 
spectral functions is a 
clear manifestation 
that correlations are 
indeed important. For 
smaller density/spin, 
for example $\rho = 0.1$ 
(Fig. 8), we find that the 
spectral functions are 
almost single peaks, 
or free-like quasi-particles. 
We have 
compared our self-consistent 
calculation with 
a low order approximation 
which consists 
in replacing the internal 
one-particle 
Green functions 
by free ones. The effects 
of the latter 
approximation are evident: the 
spectral functions 
are almost delta functions 
and the self-energy 
shows a wide gap around 
the chemical 
potential, which signals 
that this 
approximation is 
mean-field-like. We 
add that besides 
the good features 
present in the FLEX 
approximation\cite{Dahm}, 
we have shed some light on another 
aspect of it, i.e., the 
dependence of the one-particle 
properties on the two-particle 
Green functions due to the 
self-consistency of the 
FLEX equations.  
We mention the work of 
Nakamura, Moriya and 
Ueda\cite{22} and references
therein where they point 
out the role of both 
the low and high frequency 
behavior of the 
spin fluctuations in the 
superconducting phase. 
For densities around 
half-filling correlations 
start to build in 
giving rise to three peaks 
in the energy 
spectrum. 
We argue that this analysis 
can be described in 
a generalized scheme of three-pole Ansatz 
for the spectral 
function\cite{Nolting}. 
Work along these line 
is in progress\cite{JJMA} in order to really 
resolve the three peaks in the one-particle 
spectral function. We should add that three 
peaks in the $A(\vec{k},\omega)$ is equivalent 
to two peaks in the imaginary part of the 
self-energy. In this work, we have also 
discussed the differences of FLEX with respect 
to the $T$-Matrix approximation. 

\section{Acknowlegments.}
We would like to thank Brazilian 
Agency CNPq (Project 
No. 300705/95-96), CONDES-LUZ 
and CONICIT 
(Project No. F-139). Very useful 
discussions with Profs. 
M.S. Figueira, 
E.V. Anda, M.A. Continentino, 
A.M. Rodero and H. Beck  
are fully acknowledged. 
We appreciate Prof. 
G. Mart\'{\i}nez for 
calling our attention 
to references 
\cite{16},\cite{17}. 
We thank  
Prof. Mar\'{\i}a Dolores 
Garc\'{\i}a 
Gonz\'alez for reading 
the manuscript. 
%\end{multicols}
%\begin{multicols}{2}

%\end{multicols}

\vspace{.5cm}
%\newpage
%\begin{multicols}{2}
%\narrowtext
\begin{center}
{\Large Figures.}
\end{center}

\begin{figure}
\caption{ 
The density/spin, 
$\rho$ as function 
of chemical potential, 
$\mu$ for the full 
self-consistent 
FLEXC equations as 
given in 
Eqs. (\protect{\ref{Dyson}},\protect{\ref{rho}}).
We also show the 
results for a low 
order approximation where we 
substitute the internal 
one-particle 
Green functions by 
free Green 
functions.  The 
parameters used here 
are $U/t = 4.0,~T/t = 0.03$.}
\end{figure}

\begin{figure}
\caption{
The diagonal one-particle 
spectral function, 
$A(n(\pi,\pi),\omega)$ vs
$\omega$ for 
different momenta 
along the diagonal of the 
Brillouin zone $(\vec{k} = 
(n,n)\pi/16)$ for 
$U/t = 4.0$, 
$T/t = 0.03$.  
We have used an external
damping of 
$\delta/t = 0.01$, 
$32 \times 32$ 
points in the 
Brillouin zone and 
1024 Matsubara 
frequencies. After 
self-consistent 
calculation of the coupled
non-linear equations, 
we get $\mu/t = 
- 0.9558$.  We have 
runned our source 
code in single 
precision requiring 
$43~MB$ of RAM 
memory. Each iteration 
takes 3.2 
minutes of CPU time 
in a Pentium 166.}
\end{figure}

\begin{figure}
\caption{
The diagonal one-particle 
spectral function, 
$A(n(\pi,\pi),\omega)$ vs
$\omega$ for different 
momenta along 
the diagonal of the
Brillouin zone $(\vec{k} 
= (n,n)\pi/16)$ for the 
low order approach. 
Same parameters 
as in Fig. \ 2. Here we have taken 
$\delta/t = 0.0001$.}
\end{figure}

\begin{figure}
\caption{
$-Im[\Sigma(n(\pi/16,\pi/16),\omega)]$ 
vs $\omega$ for different 
momenta along the
diagonal of the Brillouin zone 
(${\vec{k}} = (n,n)\pi/16$).  
Same parameters as in
Fig.\ 2.}
\end{figure}

\begin{figure}
\caption{
(a) $-Im[\Sigma(n(\pi/16,\pi/16),\omega)]$ 
vs $\omega$ for 
different momenta along the
diagonal of the 
Brillouin zone 
(${\vec{k}} = 
(n,n)\pi/16$) for the low order 
approximation. 
Same parameters as in
Fig. \ 2. $\delta/t = 0.0001$. (b) $-Im[\Sigma(n(\pi/16,o,\omega)]$ 
with the same parameters.}
\end{figure}

\begin{figure}
\caption{
$Im[V_{p-h}(m(\pi/16,\pi/16),\omega)]$ 
vs $\omega$ for different 
momenta along the
diagonal of the Brillouin zone 
($\vec{q} = (m,m)\pi/16$). 
Same parameters as in
Fig. \ 2. To simplify a little bit the notation, 
we have identified the effective potential, 
$V_{p-h}(m(\pi/16,\pi/16),\omega)]$ by 
$T(m(\pi/16,\pi/16),\omega)]$. The same 
is done in the next figure.}
\end{figure}

\begin{figure}
\caption{
$Re[V_{p-h}(m(\pi/16,\pi/16),\omega)]$ 
vs $\omega$ for different 
momenta along the
diagonal of the Brillouin zone 
($\vec{q} = (m,m)\pi/16$). 
Same parameters as in
Fig. \ 2.}
\end{figure}
%\end{multicols}

\begin{figure}
\caption{ 
The diagonal one-particle 
spectral function, 
$A(n(\pi,\pi),\omega)$ vs
$\omega$ for different momenta 
along the diagonal of the
Brillouin zone $(\vec{k} 
= (n,n)\pi/16)$. 
Here the density/spin is 0.1 
and $\mu = -3.17t$.}
\end{figure}

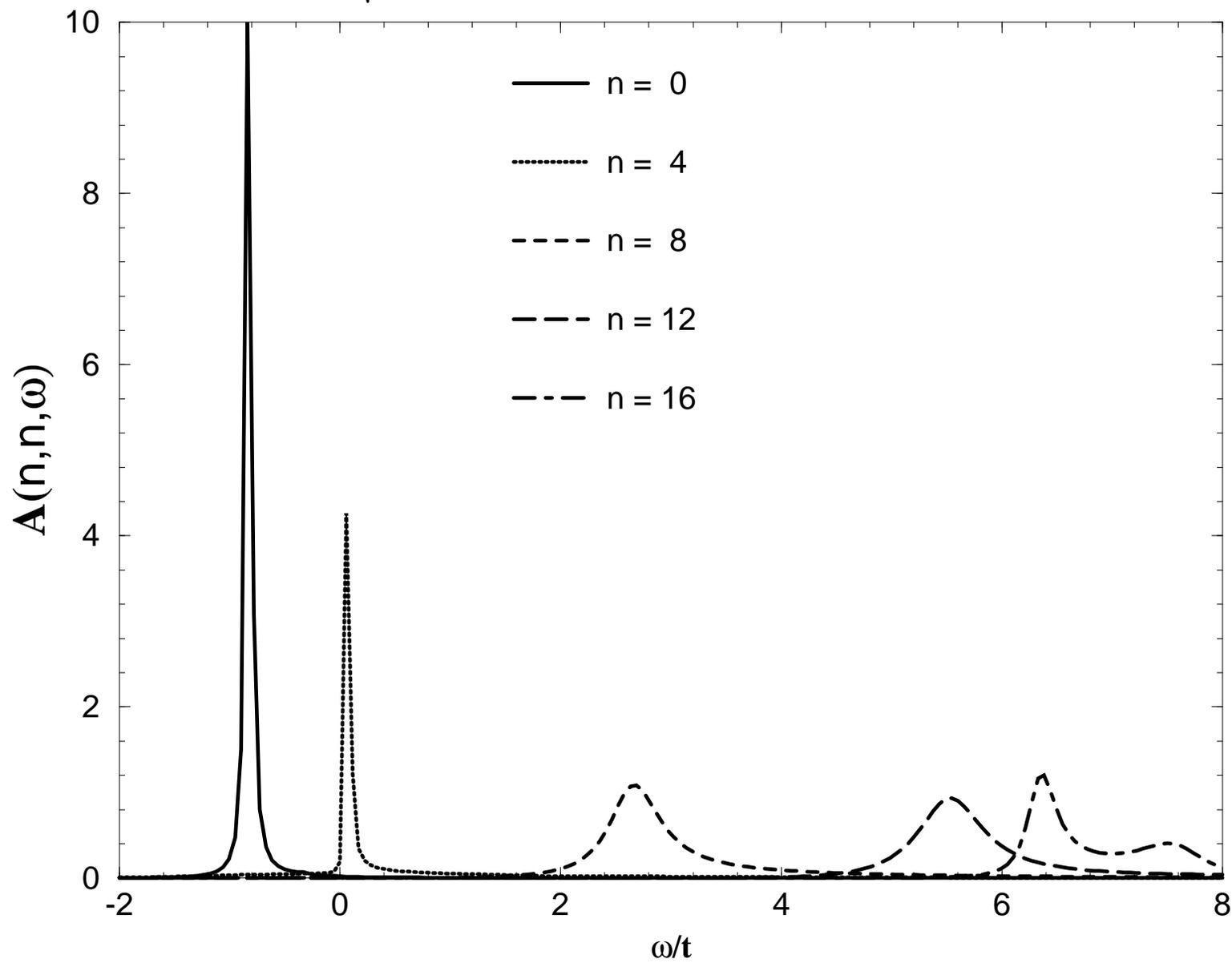
\end{document}